\documentclass[12pt]{spieman}  
\usepackage{amsmath,amsfonts,amssymb}
\usepackage{graphicx}
\usepackage{setspace}
\usepackage{tocloft}
\usepackage{lineno}

\title{Real-time station monitor and stationtest pipelines for LOFAR~2.0}

\author[a,*]{Jun Wang}
\author[b]{M.~J. Norden}
\author[b]{P. Donker}
\affil[a]{Ruhr-Universit\"at Bochum, Fakult\"at f\"ur Physik und Astronomie, Astronomisches Institut (AIRUB), 44801 Bochum, Germany}
\affil[b]{Netherlands Institute for Radio Astronomy (ASTRON), Postbus 2, 7990 AA Dwingeloo, The Netherlands}

\cftpagenumbersoff{figure}
\cftpagenumbersoff{table} 

\begin{document} 

\maketitle

\begin{abstract}
LOFAR is a low-frequency array distributed across several European countries. Each LOFAR station contains thousands of antennas and associated electronics, making monitoring and thorough testing of those components essential to ensuring station reliability. This paper discusses various anomalies that may arise in LOFAR antennas, tile elements, modems, and summators. We also introduce two diagnostic pipelines designed to detect these anomalies: a real-time station monitoring system and an offline stationtest system. These pipelines provide valuable insights into the operational status of each antenna, issuing alerts to minimize observational disruptions while maximizing station uptime, reliability, and sensitivity. By enhancing the efficiency and stability of LOFAR stations, they also serve as a foundation for future large-scale arrays like SKA-Low. The experience gained from their development and deployment will contribute to the construction and maintenance of SKA-Low, improving monitoring and diagnostic capabilities for large-scale antenna networks. Ultimately, these systems play a crucial role in ensuring continuous observations and maintaining data integrity.
\end{abstract}


{\noindent \footnotesize\textbf{*}Jun Wang,  \linkable{jun.wang.ucas@gmail.com} }

\begin{spacing}{2}   

\section{Introduction}
\label{sect:intro}  
Low-frequency radio astronomy, typically covering frequencies below 300 MHz, is a crucial part of the radio spectrum for observing the universe. It facilitates the study of key astrophysical phenomena, such as the early stages of galaxy formation\cite{ghs+18}, the structure and dynamics of cosmic magnetic fields\cite{bah+13}, and transient events like pulsars and fast radio bursts\cite{sha+11, pmb+21}. Early progress in the field was driven by pioneering instruments such as Jansky's dipole array and the Reber radio telescope, both of which operated within the low-frequency range. A significant milestone was achieved with the development of the Interplanetary Scintillation Array, which enabled the first detections of pulsars thanks to its high-time resolution capabilities\cite{hbp+68}. More recently, telescopes like the Long Wavelength Array (LWA\cite{ecc+09}) and the Giant Metrewave Radio Telescope (GMRT\cite{swa91}) have further advanced the field, providing higher resolution and improved sensitivity. Building on these foundational efforts, European scientists have developed the Low-Frequency Array (LOFAR\cite{vwg+13}), a next-generation instrument that utilizes cutting-edge digital signal processing to achieve unprecedented sensitivity and angular resolution, thus opening new frontiers in low-frequency radio astronomy.

As the world's largest and most sensitive radio telescope operating at low radio frequencies, LOFAR (covering a frequency range of 10 to 240~MHz) is a distributed phased array telescope composed of multiple antenna stations across Europe. These stations are software-driven and supported by powerful computing and extensive data storage in several distributed data centers\cite{mms+24}. Together, they form a unified, agile, and highly capable system for observing and data processing. LOFAR, leveraging its 2000~km maximum baseline and high-sensitivity antennas, offers sensitivity and angular resolution at least 30 times greater than previous low-frequency telescopes, achieving angular resolutions ranging from $0.5^\circ$ to sub-arcsecond scales. Additionally, it can observe simultaneously in up to 488 beams, providing exceptional imaging capabilities for large-scale sky surveys\cite{vwg+13}.

\begin{figure}[ht!]
    \centering
    \includegraphics[width=0.9\columnwidth]{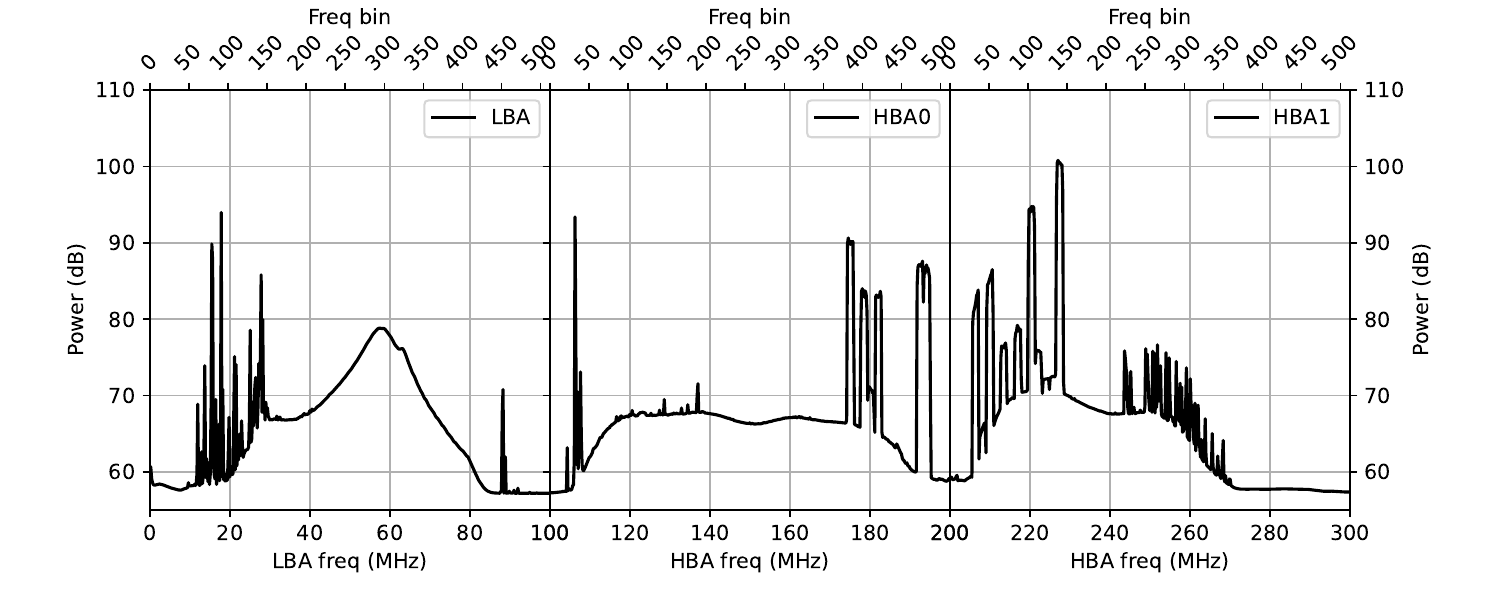}
    \caption[Simultaneous power spectra of LBA and HBA antennas, taken from station CS001.]{Simultaneous power spectra of LBA and HBA antennas, taken from station CS001. From left to right, the spectral plots of LBA and HBA (which can be divided into two sub-arrays, HBA0 and HBA1, referred to as `ears') at different sampling frequencies are shown.}
    \label{fig:l2ts_powerspectral}
\end{figure}

As of 2024, the LOFAR network comprises 52 stations across 8 European countries. Among these, 38 Dutch stations are located in the Netherlands, while the remaining 14 international stations distributed as follows: 6 in Germany, 3 in Poland, and one each in France, Sweden, the United Kingdom, Ireland, and Latvia. In addition, two new stations, located in Italy and Bulgaria, are currently in the planning phase and will be constructed as part of the LOFAR 2.0 upgrade project. Each station is equipped with 96 Low-Band Antennas (LBA) for the 10-90 MHz frequency range and either 48 or 96 High-Band Antenna (HBA) tiles for the 110-240 MHz range. Figure~\ref{fig:l2ts_powerspectral} shows the simultaneous power spectral density of both LBA and HBA, a new feature enabled by the LOFAR 2.0 hardware upgrade in the stations.

Each HBA tile consists of an array of 16 dual-polarized bow-tie antenna elements arranged in a 4$\times$4 grid with 1.25~m spacing between the dipoles, forming a 5$\times$5~m structure. Each tile is equipped with an analog radio frequency beamformer that combines the signals from the 16 individual antenna elements in phase to create a single ``tile beam'' focused on a specific direction in the sky. The pointing of the HBA tile beam is controlled using a five-bit delay line structure, allowing for precise delay adjustments in steps of 0.5, 1, 2, 4, and 8~ns. The maximum achievable delay is 15.5~ns, providing a time resolution of 0.5~ns, which is sufficient to correct incoming signals over a distance of 4.65~m.

The control of these internal delays within the HBA tile is managed by a modem. Signals from the 16 HBA elements within a tile are routed through two 16-to-1 summators. The power summator (X-Pol) is responsible for biasing all elements within the tile, while the communication summator (Y-Pol) handles the transmission of delay settings to the 16 HBA front-end elements.

In addition to the antennas, each station also includes a cabinet that houses various electronic components essential for receiving, pre-processing, and transmitting astronomical data. This cabinet contains a range of equipment, including receiver units (RCUs), analog-to-digital converters (ADCs), digital signal processing units, power converters, local control computer and switches.

Like single-dish radio telescopes, which require precise monitoring of position and orientation\cite{zdxb19}, antenna array telescopes also demand sophisticated and highly accurate monitoring systems. Due to the vast number of components and the complexity of antenna arrays, continuous monitoring of each element's status is critical for ensuring smooth operation. This real-time monitoring helps detect potential issues early, preventing any degradation in system performance. With this purpose, Ref.~\citenum{ba07} initially introduced a Ganglia monitoring system for telescope arrays. However, with advancements in the Tango Controls framework\cite{gtp+03}, many telescopes and antenna arrays now employ integrated architectural systems that provide centralized monitoring and control\cite{pdr+17, lmff24}.

Monitoring data is utilized downstream from the collection point to offer a passive, real-time summary of the system's status\cite{ba07}. By tracking the operational status of each component, engineers can swiftly identify and address malfunctions, shut down and repair faulty elements, maintain optimal system performance, and, most importantly, ensure the accuracy and reliability of the scientific data collected by the array.

In this paper, we present an overview and reference description of the LOFAR 2.0 real-time station monitor (RTSM) pipeline, as well as the stationtest routine. The structure of this paper is outlined as follows: Section~\ref{sec:lofar2} briefly introduces the ongoing LOFAR 2.0 project and its upgrade targets. The anomalies present in the LOFAR system are explained in Section~\ref{sec:anomalies}. Following that, in Section~\ref{sec:rtsm}, we describe the development and results of both the RTSM and stationtest. Later in Section~\ref{sec:system}, we present system implementation and technical overview of the system. Finally, we summarize our conclusions and envision the station monitoring and testing pipeline for SKA in Section~\ref{sec:summary}.

\section{LOFAR 2.0} \label{sec:lofar2}
LOFAR 2.0 is an upgraded version of the original LOFAR radio telescope, designed to enhance the performance, capabilities, and scientific reach of the existing system (hereafter referred to as ``LOFAR 1.0''\cite{vwg+13}). With advances in technology and the increasing complexity of data, LOFAR 1.0 has encountered challenges in meeting modern astronomical demands. The LOFAR 2.0 project addresses these limitations by offering improved sensitivity, greater bandwidth, enhanced computational power and a more precise clock system, ensuring that LOFAR remains at the forefront of astronomical research into the SKA era.

One of the most significant changes in LOFAR 2.0 is the comprehensive hardware upgrade, which includes improvements to both front-end and back-end systems. The front-end now features redesigned low-noise amplifiers (LNAs), increasing antenna sensitivity and making them more efficient at detecting faint radio signals from space. In addition, the DANTE project (Development of an Advanced HBA Front-end) aims to enhance the capabilities of LOFAR's HBA system by developing a new HBA front-end electronics (HBA-FE) board and designing an advanced HBA tile summator-beamformer. This will enable the generation of two fully independent beams per tile, significantly improving the system's observational capacity. A key innovation is the introduction of UniBoard2, an advanced processing board that dramatically improves data handling capabilities\cite{ssg+19}. Additionally, the system now includes COBALT 2.0, the second version of the Correlator and Beam-former, enabling more powerful and flexible data processing\cite{skh+24}. These upgrades enable faster and more efficient processing of incoming signals, handling data volumes 3-6 times larger.

Furthermore, new monitoring units have been integrated into the system, offering more precise control and diagnostics of the telescope's operations. Collectively, these upgrades enable LOFAR 2.0 to operate with increased sensitivity and energy efficiency, while expanding its ability to observe a broader range of radio frequencies simultaneously.

In terms of software, LOFAR 2.0 introduces several critical upgrades. The Telescope Monitor Specification System (TMSS) provides more comprehensive oversight of the entire array, enhancing the operational management of the telescope. Dutch LOFAR stations now benefit from improved clock synchronization via the White Rabbit\cite{msw+09} protocol, which ensures nanosecond precision across the array for more accurate data correlation\cite{wbm+24}. To improve real-time system monitoring and control, the Grafana platform has been integrated, offering a user-friendly dashboard for system diagnostics\cite{jms+21}. The backend for Grafana relies on Prometheus for monitoring metrics\cite{lmff24, jms+21} and Loki for log aggregation, both integrated with the Local Control Unit (LCU) of each station. Metrics are collected from Tango devices and other components, while logs are parsed and forwarded by Vector before being aggregated in Loki. Additionally, LOFAR 2.0 incorporates Jupyter Notebooks, enabling operators and engineers to efficiently control stations and analyze raw data. Together, these software enhancements, combined with the hardware upgrades, greatly expand LOFAR 2.0's data processing capacity, ensuring it can handle larger datasets with greater precision and flexibility.

LOFAR 2.0 also focuses on improving operational efficiency and extending the telescope's lifespan\cite{oni+24}. LOFAR 1.0 required significant manual intervention for calibration, maintenance, and monitoring, leading to increased downtime and operational costs. In contrast, LOFAR 2.0 automates key processes, reducing the need for human oversight and improving system reliability. This shift not only enhances scientific productivity but also extends the telescope's operational life by replacing outdated components and incorporating more durable and sustainable technologies. Furthermore, under the newly established LOFAR European Research Infrastructure Consortium, the integration of additional international stations, such as those planned in Italy and Bulgaria, strengthens European collaboration, further boosting the array's sensitivity.

\section{LOFAR antenna anomalies} \label{sec:anomalies}
LOFAR's complexity stems not only from its vast network of antennas spread across multiple stations in different countries but also from the fact that each station contains hundreds of LBA antennas and HBA tiles, each tile comprising multiple elements. To achieve this, LOFAR 2.0 employs two key diagnostic mechanisms: an universal real-time station monitor (RTSM) and a dedicated stationtest for HBA elements. These tools are essential for detecting and diagnosing issues across the system, from individual antennas and tiles to the elements within each tile.

RTSM is designed to continuously assess the operational status of each antenna in the station, from the individual antennas to the tiles. Given the scale of LOFAR, any failure in a single antenna or tile can affect overall data quality and scientific output. RTSM provides timely alerts on system health, allowing operators to quickly identify and address issues such as hardware failures, synchronization problems, or signal anomalies. Early detection minimizes downtime, optimizes performance, and ensures that scientific observations are not compromised. This real-time capability is critical for maintaining LOFAR's reliability, especially during long observation periods or while monitoring transient astronomical events, where rapid response and highly reliable electronics are essential.

Stationtest, by contrast, is a dedicated diagnostic tool used to perform in-depth tests on station components, particularly the elements within each tile. While RTSM focuses on detecting real-time anomalies during operations, stationtest systematically verifies the performance of each component under controlled offline conditions, including tests that cannot be run during observations. This is particularly important, as minor issues of elements within tiles can cumulatively affect the array's overall performance. Stationtest ensures that each part of the antenna system functions correctly, identifying issues before they escalate or impact scientific data collection. By running these tests regularly, the LOFAR team maintains high data integrity, accurately calibrates the system, and prevents the gradual accumulation of faults that might otherwise go unnoticed.

The necessity of these testing tools is driven by LOFAR's need for precision and reliability in its observations. With thousands of antennas working together, even minor issues can affect the coherence and quality of the signals being collected. RTSM helps maintain operational efficiency by providing immediate alerts, while stationtest ensures long-term stability and accuracy through thorough diagnostics. Together, these tools form a comprehensive monitoring and testing framework, ensuring LOFAR operates at peak performance and continues to deliver the high-quality data required for groundbreaking astronomical research.

LOFAR antennas, particularly the LBA systems, can encounter a variety of issues that affect their performance and the quality of the data they collect. Common anomalies include down, short, flat, spurious signals, oscillation, noise (jitter, high or low levels), RF-power issues, cable reflections, and failures in the HBA front-end element, modem and summator. Below is a detailed explanation of each of these issues:

\begin{description}
  \item[\textbf{1. Down} (antenna collapse or leaning)]
The term ``down'' refers to a situation where an antenna is no longer functioning properly, often due to physical damage or electrical failure. Since LBA is secured to the ground using polyester rope and rubber spring, external factors such as animals chewing on the cords and harsh weather conditions (e.g., strong winds or snow), the antennas can sometimes become unstable and fall over. This can result in interruptions in signal reception and require regular maintenance to ensure the antennas remain in their proper position and function effectively.

\item[\textbf{2. Short} (short circuit)]
A ``short'' circuit occurs when there is unintended electrical contact between conductors, causing a disruption in signal transmission. This could happen in the antenna itself, or due to insulation failure in the cables or connectors, leading to an electrical short. As a result, the antenna may totally lose its ability to receive astronomical data, significantly degrading performance of the antennas.

 \item[\textbf{3. Flat }(flat signal response)]
The ``flat'' condition indicates that the antenna is producing a signal response that lacks the expected variations over frequency. This often points to an issue in the amplification or filtering systems, where the antenna's frequency response becomes uniform, diminishing its capability to accurately observe signals at the resonant frequency. It can also suggest a problem with signal processing, leading to poor or unusable data.

 \item[\textbf{4. Spurious} (unwanted signals or spurious emissions)]
``Spurious'' signals refer to unwanted or unexpected signals detected by the antenna, often occurring in frequency bands that are typically free of interference, where the associated radio sources do not fall within the range of usual interference sources. These may originate from external sources such as nearby electronic devices, lightning, or man-made radio frequency interference (RFI). Spurious emissions can contaminate astronomical data, making it difficult to detect and analyze true astronomical signals. Identifying and mitigating spurious signals is essential for maintaining data integrity.

 \item[\textbf{5. Oscillation} (self-induced signal oscillation)]
Oscillation refers to the situation where an antenna's electronics, such as amplifiers or circuits, begin generating uncontrolled repetitive signals. This can be caused by faulty components or design issues within the antenna system. In some cases, oscillation creates feedback loops that can exacerbate issues, pushing certain components beyond their operational limits. This can lead to faster wear or sudden breakdowns. While most oscillation do not directly damaging in all cases, oscillation distorts the signals received by the antenna, introducing false data or noise, especially coherent beamformed observations, such as those for pulsars, can greatly suffer from this.

 \item[\textbf{6. Noise} (high or low noise levels, and jitter)]
Noise levels refer to the amount of unwanted signal interference detected by the antenna. High noise levels can obscure faint astronomical signals, making it harder to distinguish real data. This excess noise may come from thermal emissions, electrical interference, or environmental factors. Low noise levels, while seemingly positive, could indicate a malfunction in the antenna's signal detection systems, preventing it from capturing enough signal to produce useful data. Jitter is the deviation in the signal's phase or timing, which can cause the arrival time of signals to fluctuate unpredictably. It can manifest as distortions or irregularities in the frequency domain. Instead of a clean, stable signal, the presence of jitter can cause broadening of spectral peaks and phase noise. In overall, the goal of the noise check test is to maintain an optimal signal-to-noise ratio for accurate measurements.

 \item[\textbf{7. RF-Power} (high or low radio frequency power levels)] RF-power refers to the strength or intensity of the signal being received by the antenna. If the RF-power is too high, it can lead to signal distortion or overload the system's electronics, causing inaccurate data. On the other hand, if RF-power is too low, the antenna might not be able to detect weaker signals, resulting in poor observational data. Anomalous RF-power levels usually indicate problems with either the external environment (e.g., interference) or the antenna's internal components, such as amplifiers.

 \item[\textbf{8. Cable Reflection} (signal reflection in cables)]
Cable reflection occurs when part of the signal being transmitted through the antenna's cables is reflected back toward the source. This usually happens due to impedance mismatches in the cables or connectors, often caused by poor connections or damaged cables. Signal reflection results in loss of signal strength and quality, as well as potential interference with the incoming signal, reducing the overall effectiveness of the antenna.

 \item[\textbf{9. HBA-specific modem error}]
The modem in a tile are responsible for managing the signal delays across the individual elements of the tile to ensure proper synchronization for tile analog beamforming. Ensuring accurate delays is critical for the proper alignment of signals from multiple elements. Incorrect element delays can lead to signal desynchronization, reducing the coherence of the signals, which in turn affects the accuracy of the analog beamforming data.

 \item[\textbf{10. HBA-specific summator noise}]
The summator is responsible for combining signals from different elements of the HBA tiles to create a coherent output that can be processed by the system. The summator noise is caused by the direct current conversion on the P-summator. When the shield is not proper closed or connection around the connectors is open, the electromagnetic interference from the conversion can leak to the antenna elements.

\end{description}

\begin{figure}[ht!]
    \centering
    \includegraphics[width=0.95\columnwidth]{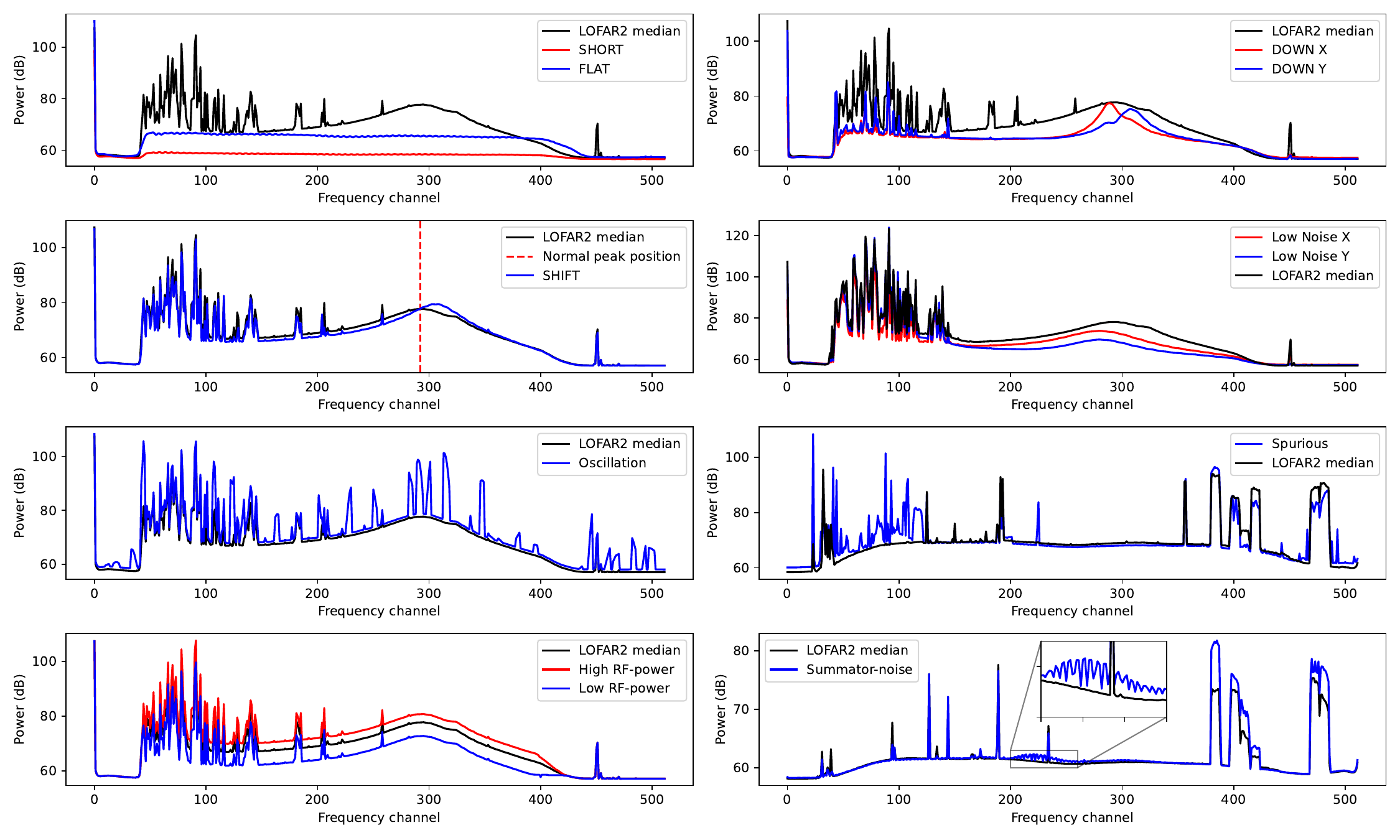}
    \caption[Power spectral behaviors of several anomalies.]{Power spectral behaviors of several anomalies. In each panel, the median power spectrum of the LBA or HBA antennas (represented by black lines) is displayed along with the power spectra corresponding to anomalies (shown as solid red or blue lines).}
    \label{fig:anomalies_powerspectral}
\end{figure}

In Figure~\ref{fig:anomalies_powerspectral}, we presented power spectra of each anomalies. These anomalies underscore the complexity involved in maintaining the LOFAR antenna array and highlight the critical importance of continuous monitoring and testing. Timely detection and resolution of these issues are essential to ensuring the array operates efficiently and consistently delivers high-quality astronomical data.

\section{LOFAR RTSM and Stationtest} \label{sec:rtsm}
To mitigate the impact of antenna anomalies on LOFAR's performance and the quality of observation data, we have developed two monitoring systems: the real-time station monitoring (RTSM) and stationtest. The RTSM operates concurrently with observations, reporting any antenna anomalies detected in real time. Additionally, by leveraging LOFAR 2.0's live statistics streaming capability, the RTSM can also analyze antenna status when the antennas are active but not in use for observations, without interfering with other testing operations.

In contrast, the stationtest system is designed to assess all HBA tile components within a station, including elements, modems, and summators. These tests involve simultaneous adjustments to modems and summators, as well as frequent toggling of elements on and off across all tiles in the station. As a result, stationtest cannot be performed on individual tiles while others continue observing. Instead, it requires suspending observations for the entire station and allocating dedicated telescope time to complete these diagnostics. The stationtest is run on all station tiles simultaneously, which effectively suspends observations for the duration of the tests. LOFAR stations are allocated fixed time slots for stationtest, typically scheduled once a week, to conduct thorough checks of tile-elements. Each stationtest session lasts 1-2 hours, depending on the number of cycles performed during the run.

\subsection{RTSM for LBA and Tile-based HBA}
In LOFAR 2.0, as in LOFAR 1.0 as well, three types of statistical data are automatically generated to monitor system performance: subband statistics (SST), beamlet statistics (BST), and crosslet statistics (XST). SST track the power levels of each antenna polarization and specific frequency subbands, enabling detailed monitoring of frequency-dependent behavior and detecting potential interference or signal degradation. This is critical for maintaining signal quality across LOFAR's broad frequency range, as irregularities in specific subbands can affect overall data quality. SST data is extensively used in both the RTSM and stationtest systems to detect anomalies in antenna performance.

BST, on the other hand, monitor beamlets formed when signals from multiple antennas are combined and steered towards specific celestial targets. By tracking signal integrity within each beamlet, BST ensure that the beamforming process functions correctly, allowing for precise observations of specific regions of the sky. Lastly, XST provide cross-correlation information between antenna pairs, which is essential for interferometry. XST verify that signals from separate antennas are properly synchronized, ensuring high-resolution data and accurate imaging.

The RTSM primarily performs real-time diagnostics on the LBA antennas and HBA tiles at a LOFAR station. For LBA antennas, it executes tests 1-8, as outlined in Section~\ref{sec:anomalies}, while for HBA tiles, it focuses mainly on tests 4-10. RTSM analyzes spectral data, using SST primarily to assess the state of each antenna.

RTSM runs directly on the station LCU, with a typical test cycle for both LBA and HBA antennas taking approximately 60 seconds. However, the cycle interval can be adjusted based on system processing requirements to reduce the demand on computing resources. Currently, RTSM relies on log files for storing processing history. Additionally, spectrum data from antennas exhibiting abnormal behavior are saved in a dedicated CSV file on the station's LCU, allowing for more detailed post-analysis of malfunctioning antennas.

\begin{figure}[htbp]
    \centering
    \includegraphics[width=0.85\columnwidth]{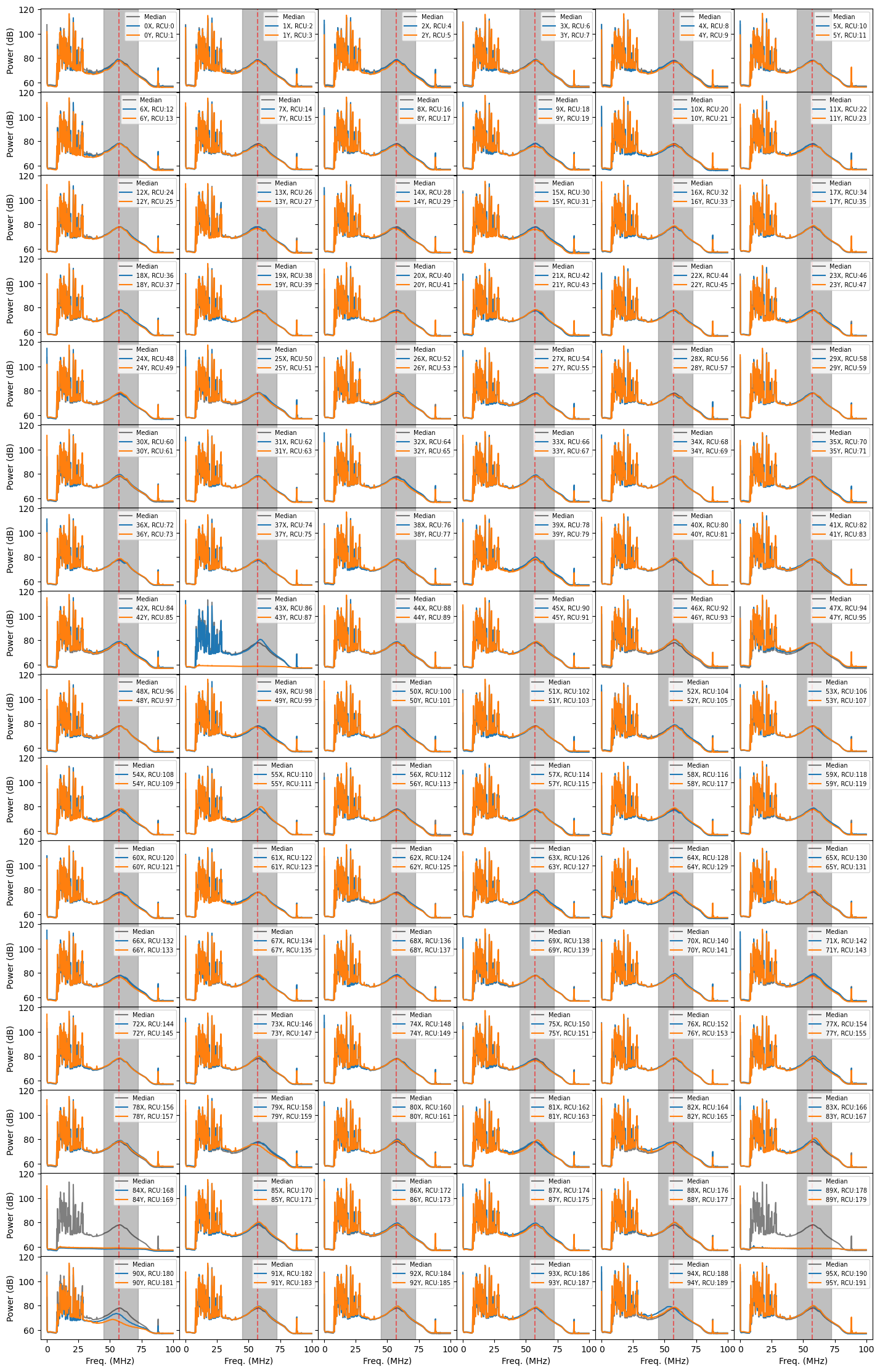}
    \caption[An image of the LBA frequency spectrum, generated by the RTSM.]{An image of the LBA frequency spectrum, generated by the RTSM. From the spectrum, we can see that antennas 43Y, 84, and 89 are short, antenna 90 has low noise, and 94X is shifted.}
    \label{fig:lba_powerspectral}
\end{figure}

In Figure~\ref{fig:lba_powerspectral}, the power spectrum from the LBA of CS001 is shown. The gray curve in each subplot represents the median data across all antennas and polarizations, while the blue and orange curves correspond to the X and Y polarizations of the same antenna, respectively. The dashed orange line marks the resonant frequency of the LOFAR LBA, which is around 58~MHz. The shaded area highlights the less contaminated frequency range primarily utilized by the RTSM.

\subsection{Modem Check and HBA Element-based Stationtest}
To test the elements and electronics within an HBA tile, a targeted station test is required. Typically, the modem, summator, and the 16 elements in each tile must be regularly tested to ensure system reliability and stability. During the stationtest, the modem is tested first, followed by individual elements and the summators. We conduct stationtest once a week, each lasting 1-2 hours. Since the environments of each station vary greatly, the specific error rate will vary depending on station differences and repair cycles.

Each element in an HBA tile is assigned a delay to ensure that signals from all elements are correctly timed for summation and processing. In the modem test, all elements and their corresponding LNAs remain activated. Different delay times are applied to each element via the modem. After setting these delays, the modem reads back the actual delay values from the hardware and compares them with the originally set values to detect any discrepancies. The delays are adjusted in 0.5~ns increments, ranging from 0 to 15.5~ns, controlled by a 5-bit delay unit.

\begin{figure}[htbp]
    \centering
    \includegraphics[width=0.9\columnwidth]{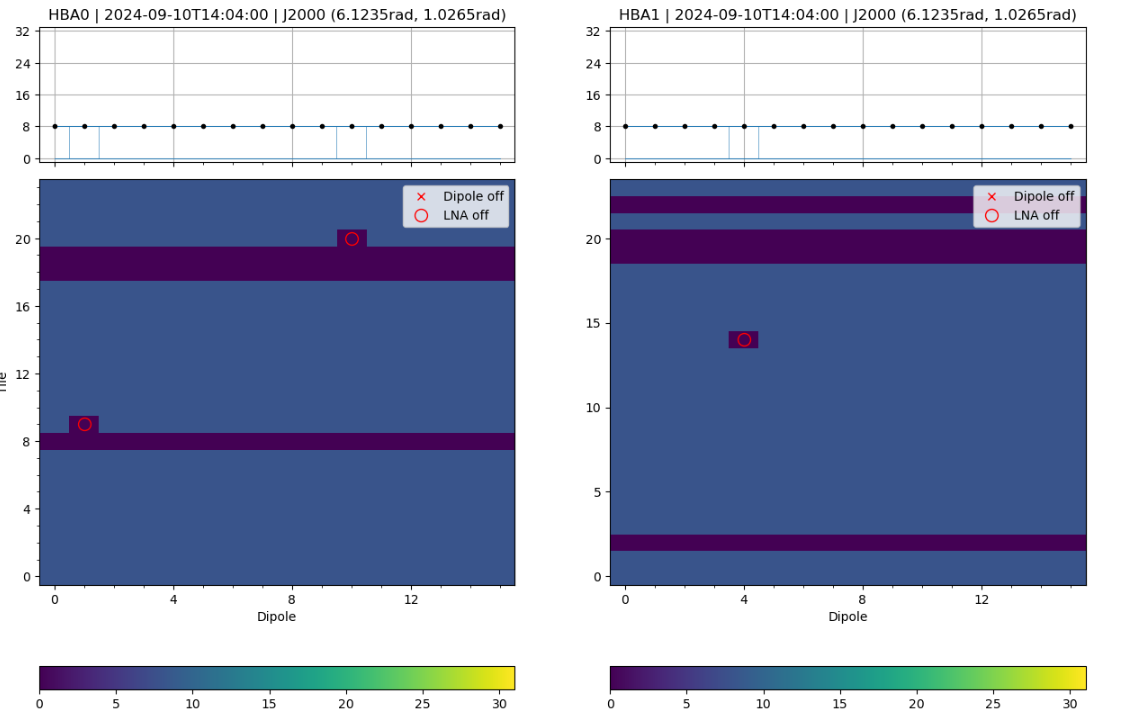}
    \caption[Example of HBA modem test obtained from the stationtest, with HBA delay factor set to 8, which correspond a HBA delay of 4~ns.]{Example of HBA modem test obtained from the stationtest, with HBA delay factor set to 8, which correspond a HBA delay of 4~ns. The long dark band in the figure corresponds to a tile labelled as broken or beyond repair in the system. The dark region block with a red circle indicate an element communication error detected by the modem test.}
    \label{fig:l2ts_hba_modem}
\end{figure}

In Figure~\ref{fig:l2ts_hba_modem}, the testing process of an HBA modem from station CS001 is shown, with the left and right panels representing the two HBA subarrays: HBA0 and HBA1, respectively. The vertical axis of the upper plots and the color scheme of the lower plots indicate the corresponding HBA delay. In this example, the binary control unit for the HBA delay is set to 8, corresponding to a 4~ns delay. The dark bands in the lower plot indicate an HBA delay of 0~ns, corresponding to several known broken tiles in CS001 that are typically broken or beyond repair. In general, we have designed a set of antenna status markers for LOFAR 2.0. Generally, antennas are classified into four states: `OK', `BROKEN', `SUSPICIOUS', and `BEYOND REPAIR'. Antennas that are functioning normally are marked as `OK' in the system, while those that are deemed abnormal or damaged are marked as `BROKEN' or `SUSPICIOUS'. Antennas identified as difficult to repair after inspection are marked as `BEYOND REPAIR'. The cross (X) and circle (o) markers represent dipole elements and LNAs are turned off. From Figure~\ref{fig:l2ts_hba_modem}, it is clear that, aside from the tiles beyond repair, the modem is functioning normally overall. Additionally, we observe that three tiles each have one element in an abnormal state.

The HBA element test focuses on evaluating each individual element within an HBA tile. Typically, this test is conducted under two different HBA delay settings, generally at 0~ns and 15.5~ns delays. During each test, only one element and its corresponding LNA within the tile are activated, while the other 15 elements and LNAs remain turned off. The test procedure mirrors that of the LBA and is repeated for each element.

\begin{figure}[htbp]
    \centering
    \includegraphics[width=0.9\columnwidth]{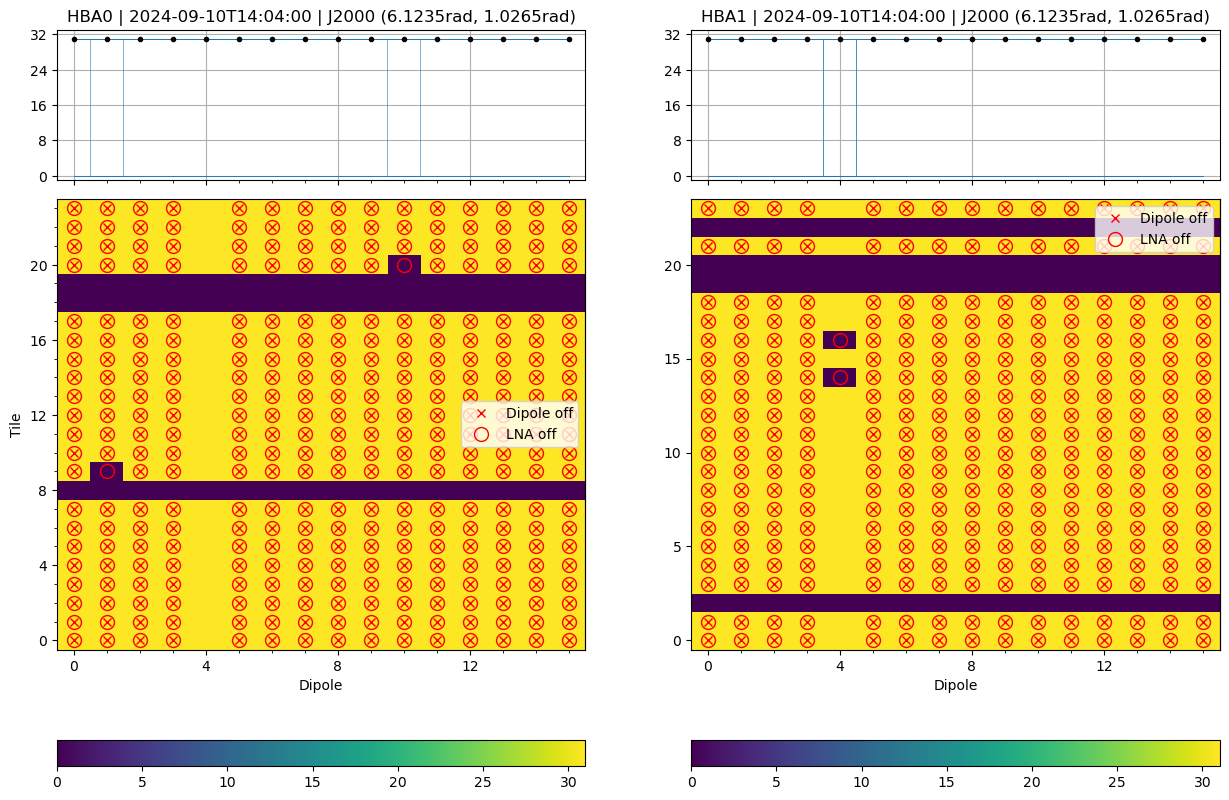}
    \caption[Example of HBA element test obtained from the stationtest, with HBA delay factor set to 31, which correspond a HBA delay of 15.5~ns.]{Example of HBA element test obtained from the stationtest, with HBA delay factor set to 31, which correspond a HBA delay of 15.5~ns. In this plot, only element 4 in each tile is set to active, while other element and correspond LNA stay inactive. However, in tile 14 and 16 of HBA1, we encounterd LNA failures. }
    \label{fig:l2ts_hba_element}
\end{figure}

As shown in Figure~\ref{fig:l2ts_hba_element}, the results are obtained with an HBA delay of 15.5~ns, where only element 4 in each tile is activated. From the figure, we observe that tile 9 and tile 20 in HBA0, along with tile 14 and tile 16 in HBA1, each have one element in an abnormal state. Further detailed testing is required to determine the exact cause of these malfunctions. Similar to the RTSM system, station tests are logged and stored as log files, with the corresponding abnormal spectrum data saved in CSV format.

\section{System Implementation and Technical Overview} \label{sec:system}
In this section, we present a detailed description on the overall software design, implementation, and operational setup of the RTSM and stationtest pipelines. By presenting these details, we aim to provide a more comprehensive understanding of how these pipelines work and their underlying architecture.

The pipelines are mainly run on the station's LCU, with Debian as the primary operating system. Each LCU is equipped with an AMD Ryzen 9 7950X 16-Core processor, 64~GB of memory and 2~TB of hard disk space for temporary storage of statistics and other data, code, etc. The pipelines mainly utilize CPU processing, and for basic operations, use very little computing resources. The user interface of the pipeline is mainly based on Jupyter Lab, providing flexibility for advanced users through configurable parameters.

The pipeline is mainly implemented in Python. Python was chosen because of its extensive ecosystem of scientific libraries and easy integration with astronomical data processing workflows, and also because in the LOFAR 2.0 control architecture, each LOFAR 2.0 station is controlled primarily through the Tango Controls API, which can be accessed very easily using the PyTango client. Therefore, each LOFAR 2.0 LCU has Jupyter Lab installed to access and control the station.

The architectural design of the pipeline emphasizes modularity to achieve scalability and maintainability. The core components include a data reading module for preprocessing raw input from the LOFAR antenna, a processing module for executing anomaly detection algorithms and spectrum analysis, and an output module for generating logs and visualizations. The design combines strong error handling and logging to ensure reliability over long periods of time. The total code is approximately 15000 lines.

The specific frequency of each type of fault depends on the location and environment of the station. Since more than 50 LOFAR stations are distributed in multiple countries with a baseline of more than 2000 km, the statistical variance varies greatly. According to our census, in the worst case, about 15\% of the antennas at LOFAR stations have faults and are turned off. Similarly, the repair time for each type of fault varies also greatly from station to station. Especially for international stations, if it is a relatively simple fault, such as an LBA antenna down, technicians at the international stations can simply repair the antenna in a shorter time. However, if it is a major component that needs to be replaced, the replacements may need to be shipped from ASTRON, so the repair time will be slightly longer. If it is a more complex fault, if ASTRON personnel need to handle it in person, it may take longer to repair. Of course, compared to repairing the antenna, it is more convenient to shut down or shield a specific antenna during the observation process, which can usually be achieved within a few minutes of observing the anomaly.

Currently, the software remains proprietary to LOFAR 2.0, but we are in the process of considering the open-source release of certain components of the LOFAR 2.0 software, including this code. If needed, the software can also be applied to antenna detection at SKA-Low with minor adjustments.

\section{Summary and Conclusion} \label{sec:summary}
This paper outlines the development of advanced tools for the RTSM and stationtest systems within the LOFAR 2.0 framework, addressing common antenna anomalies. Two dedicated programs have been developed: one for real-time monitoring of station performance and another for periodic testing of individual antenna elements. These tools were applied to station CS001, where specific anomalies, such as malfunctioning elements in HBA tiles, were identified and analyzed.

The results highlight the critical role these monitoring and testing tools play in maintaining optimal system performance by enabling rapid identification and diagnosis of issues. By continuously assessing the health of the antennas, these systems minimize downtime and ensure efficient array operation.

In parallel, the LOFAR team is developing a platform for real-time storage, visualization, and monitoring of antenna test data across all stations. This platform will provide deeper insights into the status of antennas throughout the network, improving diagnostics and enabling proactive maintenance. Future enhancements will focus on more efficient real-time data processing, automated anomaly detection, and a tighter integration of RTSM with centralized monitoring tools. These advancements are expected to reduce manual intervention and streamline telescope maintenance and performance analysis.

The techniques and methodologies developed in this work hold significant promise for future radio astronomy projects, particularly SKA-Low. Given the increased complexity and scale of SKA-Low, the real-time monitoring and testing strategies presented here will be essential in managing and maintaining such a large-scale antenna array, ensuring the system's long-term success.

\subsection*{Disclosures}
There is no conflict of interest.

\subsection* {Code, Data, and Materials Availability} 
The authors confirm that all data supporting the findings are fully available in the article. Raw data for this study are available from the corresponding author upon request.

\subsection* {Acknowledgments}
J.W. acknowledges support by the BMBF Verbundforschung under the grant 05A23PC2.


\bibliography{report}   
\bibliographystyle{spiejour}   


\vspace{2ex}\noindent\textbf{Jun Wang}  received his PhD in radio astronomy from Bielefeld University and is currently a research fellow at Astronomical Institute, Ruhr University Bochum. He has been working on the LOFAR 2.0 upgrade since 2022. His research interests include radio astronomy techniques.

\vspace{1ex}
\vspace{2ex}\noindent\textbf{Menno Norden} received his Electronic and Information degree in 1992 at the Hanzehoogeschool Groningen. Started his career at Philips Research Center in Eindhoven. From 2003 onwards working for Astron in LOFAR development program. He is currently System Engineer in the radio observatory department. His focus is on EMI protection, system issues, and monitoring.

\noindent Biographies of the other authors are not available.

\listoffigures
\listoftables

\end{spacing}
\end{document}